\documentclass[namedreferences]{solarphysics}
\usepackage[optionalrh,solaromanenum]{spr-sola-addons}
\usepackage{graphicx}                                       
\usepackage{color}                       
\usepackage{url}

\begin{document}

\begin{article}

\begin{opening}

\title{The Properties of the Tilts of Bipolar Solar Regions\\ }

\author{E.~\surname{Illarionov}$^{1}$\sep
        A.~\surname{Tlatov}$^{2}$\sep      
        D.~\surname{Sokoloff}$^{3}$
       }
       
\runningauthor{Illarionov et al.}
\runningtitle{Properties of tilt of bipolar solar regions}

\institute{
$^{1}$ Department of Mechanics and Mathematics, Moscow State University, Moscow 119991, Russia\\ 
	email: \url{illarionov.ea@gmail.com}	\\				 
$^{2}$ Kislovodsk Mountain Astronomical Station of the Pulkovo Observatory, 357700, Box-145, Kislovodsk, Russia \\
	email: \url{tlatov@mail.ru} \\
$^{3}$ Department of Physics, Moscow State University, Moscow 119992, Russia \\
	email: \url{sokoloff.dd@gmail.com}	\\	
}
             
\begin{abstract}
We investigate various properties associated with the tilt of isolated magnetic 
bipoles in magnetograms taken at the solar surface. 
We show that bipoles can be divided into two
groups which have tilts of opposite signs, and reveal
similar properties with respect to bipole area,
flux and bipolar moment. Detailed comparison of these physical quantities
shows that the dividing  point between the two types of bipoles corresponds to 
a bipole area of about 300 millionths of the solar hemisphere (MHS). The time-latitude distribution of small 
bipoles differs substantially from that for large bipoles.
Such behaviour in terms of dynamo theory may indicate that small
and large bipoles trace different components of the solar magnetic field.
The other possible viewpoint is that the difference in tilt data for 
small and large bipoles is connected with spectral helicity separation,
which results in opposite tilts for small and large bipoles. We note that 
the data available do not provide convincing reasons to prefer either
interpretation.

\end{abstract}

             
\end{opening}             

\section{Introduction}

The solar magnetic cycle is believed to be associated with dynamo 
action which occurs somewhere inside the solar convective zone. 
In turn, the solar dynamo is based on two processes. The 
differential rotation produces toroidal magnetic field ${\bf B}_T$ 
from poloidal magnetic field ${\bf B}_P$. Details of this process looks 
quite clear following the modern development of helioseismology (see {\it e.g.} the
review by \opencite{K08}). On the other hand, another process 
has to regenerate poloidal magnetic field ${\bf B}_P$ from toroidal. 
There are several solar dynamo models which suggest 
various physical mechanisms underlying this regeneration. In particular, 
the regeneration can involve sunspot formation and diffusion at 
the solar surface (Babcock-Leighton mechanism), or can be associated with 
cyclonic motions in a more or less deep layer of the convective zone only
(the Parker mechanism). However a combined action of both mechanisms looks 
possible as well (see {\it e.g.} the review by \opencite{P13}). 
The relative importance of both mechanisms for the solar cycle remains a 
topic of intensive debate. 

Of course, an observational clarification of the details of the regeneration 
process is a useful contribution to the above discussion and the tilt
angle of solar bipolar regions provides direct observational information 
for this regeneration. Indeed, the tilt data show how the 
direction between the two poles of a magnetic bipole is inclined with respect to
the solar equator. If this inclination angle differs systematically from zero, 
this would mean that poloidal field is produced from toroidal by a physical
process, which is exactly the effect under discussion. It is still not the 
whole story, and various physical mechanisms acting alone or jointly can provide
the non-vanishing tilt.  Of course, the intention of deducing the tilt from
the observational data is to clarify the physics underlying the link 
between toroidal and poloidal fields. However it is preferable not to assume 
any choice in advance, so we use below the wording '$\alpha$-effect' 
for brevity to refer to this effect.

Tilt studies originated as early as in 1919 (\opencite{Hetal19}) and resulted in 
Joy's law, simultaneously with formulation of the well-known Hale polarity law.
According to Joy's law, the average tilt is a non-vanishing quantity
antisymmetric with respect to the solar equator and growing linearly with 
$\sin \theta$ ($\theta$ is solar co-latitude). 
This result is fully in accordance with expectations from solar dynamo 
theory, and it encourages the use of the tilt data as a valuable 
observational source to constrain the governing parameters of the solar dynamo. 
The reality is however more complicated. The point is that the tilt is 
quite small (several degrees only) and rather noisy. 
Moreover, the very concept of bipoles and their identification from magnetograms
requires an algorithmic formalism in order to make comparable the results 
of independent analyses. This is probably why experts in solar dynamo theory 
did not pay attention to the tilt data for quite some time.  

The methods available for isolation of bipoles in magnetograms and the
database of the tilts grew gradually until they became convincing, at least 
for some of the dynamo community. To us, the breakthrough 
was by \inlinecite{sk12}. This paper confirms Joy's law in 
a convincing way and does not recognize any cyclic variations of the 
slope of the relation between tilt and $\sin \theta$ (see also \opencite{LU12}). 

Further investigation of the tilt data was undertaken by \inlinecite{TIS} 
who broadly confirmed the conclusions of \inlinecite{sk12} for those  
bipoles which can mainly be identified with bipolar sunspot groups.  
A time-latitude (butterfly) diagram for the tilt averaged over appropriate 
time-latitude bins obtained in \inlinecite{TIS} demonstrates that the tilt 
is indeed almost independent of the cycle phase: however some rather minor 
variations were isolated. The point however is that the analysis of 
\inlinecite{TIS} included substantially more  small bipoles than that 
of \inlinecite{sk12} and the behaviour of the small bipoles is 
almost opposite to that of the large bipoles, which correspond to sunspots. 
Recall that in \inlinecite{TIS} we referred to small bipoles
those with areas below 300 millionths of the solar hemisphere 
(MSH\footnote{1 MSH = $3.044\times10^6$ km$^2$. A round spot with area 
$S$ (in MSH) has a diameter of $d$ = $(1969\sqrt{S})$ km = 
$(0.1621 \sqrt{S})^{\circ}$ (see \opencite{Vit}).}), 
which are mostly ephemeral regions
(note that we measure in MSH an area of domains with magnetic
field exceeding the threshold level $B_{\rm min}=10~G$ isolated 
in solar magnetograms).
In particular, the tilt angle of small bipoles is antisymmetric with respect
to the solar equator, whereas the tilt of small bipoles in, say, the 
northern hemisphere is of the opposite sign to that of the large bipoles. 
\inlinecite{S13} stressed again that the analysis of \inlinecite{sk12} 
does not identify any difference between the tilts of small and large bipoles.
However this analysis is not focused on the small bipoles and the situation 
deserves further investigation and clarification.
Indeed, \inlinecite{sk12} used a substantially different approach to
bipole identification and parameters of their algorithm were optimized for
large bipoles only, thus the sample of small-scale bipoles was rather
incomplete. Moreover taking into account the higher smoothing
level applied by the authors to magnetograms, and the different type of
structures, which were recognized as bipoles, we conclude, that the sample
of small bipoles, used in \inlinecite{TIS}, cannot be compared directly
with bipoles, referred as "small" in \inlinecite{sk12}, and they are
of particular interest.

The aim of this paper is to extend the analysis of the different behaviour 
of small and large bipoles to various quantities associated with 
bipoles. It generalizes the approach of \inlinecite{sk12}, who, following the idea of the earlier research,
concentrated on the relation between tilt and latitude. The sample of 
the positions of bipoles extracted has sufficient size to 
proceed further and 
clarify a possible contribution of other factors to the tilt distribution.  

We note that the different behaviour of small and large bipoles isolated 
at least from the sample of bipoles produced by the algorithm applied does not 
seem to represent fundamental problems for dynamo theory. 
In particular, an assumption that the 
bipoles trace the toroidal magnetic field looks straightforward for large 
bipoles at least, because they are sunspots which are considered as a tracer 
for the  large-scale magnetic field generated by solar dynamo somewhere 
in the convective shell. It might be supposed that the small bipoles represent,
for example,
poloidal magnetic field and this solves the controversy. Of course, this 
is an option only and other explanations, including even a demonstration 
that the algorithm used becomes somehow inapplicable for small bipoles,
have to be considered. 

In our opinion, such considerations have to be based on an examination of the 
scaling between various physical quantities associated with bipoles.
These could include
in a plausible way the size of bipole, {\it e.g.} its flux,          
instead of its area. This is a motivation of the research discussed here. 

Speaking broadly, we arrive at the conclusion that the distinction 
between small and large bipoles can be recognized in various physical 
quantities and confirms to some extent that small and large bipoles 
trace different magnetic field components.

\section{The Data}

We used a sample of bipoles identified by the algorithm of \inlinecite{T10}. 
The method was applied to the magnetograms from \textit{Kitt Peak Vacuum 
telescope} (KPVT) for the period 1975--2003, from the \textit{Solar and Heliospheric
Observatory Michelson Doppler Imager} (SOHO/MDI) 
(\opencite{scheal}; soi.stanford.edu/magnetic/Lev1.8/) for the period 
1996--2011 and from the \textit{Helioseismic and Magnetic Imager} (HMI)
(\opencite{scetal}) for the period 2010--2013.

We used the same parameters for recognition of bipoles as \inlinecite{TIS} 
and both data samples are identical. In particular we selected domains 
with magnetic field exceeding the threshold level $B_{\rm min}=10~G$ and 
area exceeding 50 MSH. 
The parameters applied involve a large amount
of small ephemeral regions, for which it is difficult to prove that 
each isolated region corresponds to a physical entity and we can operate with
their statistical properties only. The tests presented in this paper 
and in \inlinecite{TIS} do not show any evidence, that
they are caused by a bias in the computer algorithm (see in detail \opencite{T10}).
We stress however that an independent verification of the result by 
another algorithm looks highly desirable. Such an additional
verification is obviously out of the scope of this  paper.

For a more correct determination of bipole positions we exploited only
the central part of the solar magnetogram within 0.7 of the solar disk
radius, because projection effects near the solar limb may distort the 
result substantially.
Some bipoles may have inverse polarity and
violate the Hale polarity law. This happens only in about 5\% of cases for bipolar sunspot groups ({\it e.g.} \opencite{SK10}) 
but this quantity increases substantially in going to smaller areas.

Note that the prevalent orientation is not prescribed in advance.
The bipoles are distributed in two-year time bins and $5^\circ$ latitudinal
bins and in each bin the prevalent orientation is defined as follows.
For each of the two groups of bipoles with opposite leading
polarity we compute the Gaussian approximation to the distribution 
of their tilt angles. The group with the largest amplitude defines thus the prevalent orientation in the bin.
Normally this is just the group with the larger number of bipoles.
The obtained sample is a base for further investigations according to 
additional criteria. 

For our analysis we use all bipoles in both hemispheres. 
We combine them together in such way that final angular distribution 
has a single peak, \textit{i.e.} we subtract $180^{\circ}$ 
from the angles in the second and third quadrants, and reverse the distributions in the 
southern hemisphere. The combined sample contains tilt angles in the interval
between $-90^\circ$ to $+90^\circ$. For both hemispheres the positive sign 
indicates that the domain of leading polarity is closer to the Equator 
than the trailing domain. The negative sign means on the contrary
that the domain of trailing polarity is situated closer to the Equator. 
We use the median as a robust statistic to estimate a mean tilt,
and the t-Student criterion for $95\%$ confidence intervals.

The database used gives for the bipoles the following parameters: 
the time $t$ of observation, latitude $\theta$ of the centre of the bipole, 
the area $S$, flux $F$, distance $d$ between the poles and the tilt $\mu$.

\section{Results}

We are interested mainly in correlations between the tilt $\mu$ and  the
other parameters describing a bipole. The dependence of the orientation of bipoles 
on the solar cycle is the well-known Hale's polarity law, while the dependence 
of $\mu$ on the latitude is given by Joy's law. We recall that the 
verification of Joy's law, based on an algorithmic procedure to recognise bipolar regions, confirms the law (\opencite{sk12}; 
\opencite{LU12}; \opencite{TIS}).

\begin{figure}[h]
\begin{center}
\includegraphics[width=0.7\linewidth]{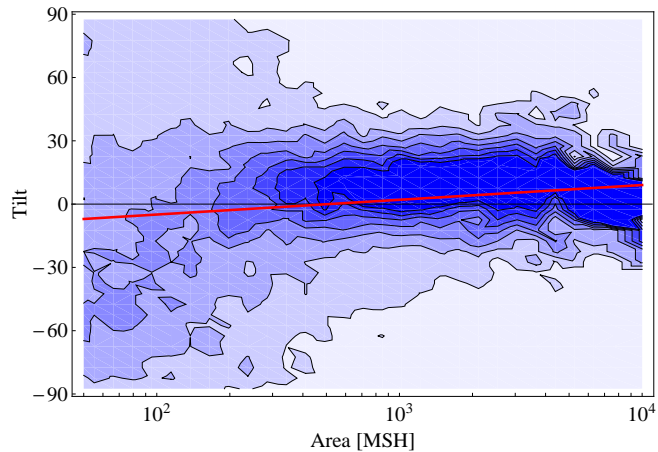} 
\end{center}
\caption{Distribution of bipoles in the combined area and tilt angle domains,
intensity of the colour indicates number of bipoles relative to the
total number of bipoles with the same areas 
(MDI data for the period 1998--2007, bipoles were selected from latitudinal zone
$|\theta|\geqslant10^{\circ}$).
The red line corresponds to the linear least-square fit.}
\label{data}
\end{figure}

Quite surprisingly \inlinecite{TIS} found that tilt substantially 
depends on the area of bipoles and that the  prevalent tilt for small bipoles 
has the opposite sign to that for large bipoles. 
This trend can be easily noticed in Figure~\ref{data}, where we show the
density of tilt angle distribution against bipole area. Indeed, for large
bipoles (we assume the dividing point between large and small bipoles
is the same as in \opencite{TIS}, \textit{i.e.} 300 MSH) we observe a pronounced peak
in the domain of positive tilts. With smaller areas the peak becomes blurred
(the distribution becomes rather non-gaussian), but the domain of increased
bipole density turns smoothly down to the domain of negative tilts.
The linear least-square fit confirms the visual trend and
intersects the line of zero tilt exactly near 300 MSH.
Difference of mean tilt signs for these two groups of bipoles is
confirmed by simple statistical test based on Student's $t$-test. 
It gives $t_{eq}=22.9$ under hypothesis that both samples have similar 
mean values and $t_{op}=0.98$ under hypothesis that mean values have
similar absolute values but opposite signs. However the noisy distribution 
for small bipoles restricts the abilities of the $t$-test in some ways.

Now, we analyse the a correlation between the 
parameters mentioned above and tilt 
on the basis of the observational data available. 
In particular, it is interesting 
to compare the correlations for large and small bipoles
in order to gain a better understanding of the physical nature of 
the difference in behaviour between these bipoles which was found in 
\inlinecite{TIS}.

\subsection{Cyclic Modulations of the Tilt}

\begin{figure}[h]
\begin{center}
\includegraphics[width=0.7\linewidth]{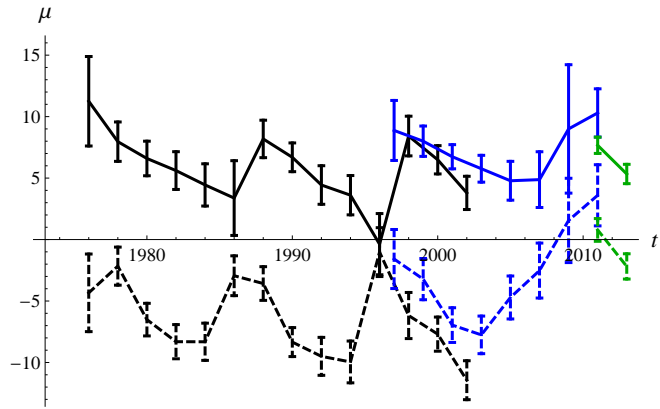} 
\end{center}
\caption{Tilt (in degrees) averaged over 2 years time bins. Solid line is for
large bipoles with areas $S>300$ MSH, dashed line is for small bipoles 
with areas $50<S<300$ MSH.
Black colour shows KPVT data, blue is for MDI data, green is for HMI data.}
\label{t_time}
\end{figure}

We start from a straightforward (and possibly not the most instructive) 
correlation property of the tilt angles, \textit{i.e.} a correlation of the tilt 
averaged over 2-year bins with the phase of the cycle. The correlation 
calculated separately for large and small bipoles is presented in 
Figure~\ref{t_time}. There are two messages from the plot. First of all, 
the KPVT data, MDI data and HMI data look to be more or less in agreement, 
at least at the epoch when the data overlap. This appears to confirm the  
self-consistency of the bipole database used.
In contrast, the large and small bipoles demonstrate opposite behaviour 
for each set of observational data.

The other message from the plot is a pronounced cyclic modulation visible 
for both types of bipoles. A remarkable feature is that in the course
of the cycle the absolute value of tilts of large bipoles decreases,
while for small bipoles we observe an increase. Strictly speaking we 
have to distinguish cycles for both types of bipoles and as it will
be shown later there are some reasons to suppose it.

An interpretation of the correlation for the large bipoles looks quite
straightforward and is consistent with the suggestion of \inlinecite{sk12}
that the slope in Joy's law is independent of the phase of the cycle. 
Indeed, Joy's law tells us that $\mu \propto \sin \theta$ where $\theta$ is 
the colatitude of the bipoles,  which decays on average with the phase.
This  results in a decay of $<\mu>$. 

\begin{figure}[h]
\begin{center}
\includegraphics[width=0.8\linewidth]{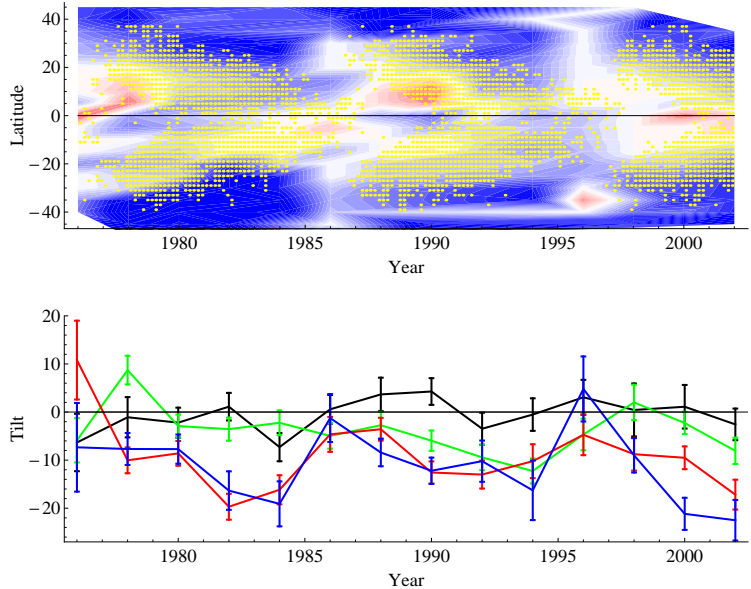} 
\end{center}
\caption{
Upper figure: time-latitude diagram for tilt according to KPVT data
for bipoles with areas $50<S<300$ MSH. 
Blue shows negative tilt, red is for positive. The yellow points
show the sunspot distribution.
Lower figure: tilt evolution for bipoles with areas $50<S<300$ MSH 
in different latitudinal zones: black represents bipoles with $|\theta|<10^\circ$,
green with $10^\circ\leqslant|\theta|<20^\circ$, red with 
$20^\circ\leqslant|\theta|<30^\circ$,
and blue with $30^\circ\leqslant|\theta|<40^\circ$.  
}
\label{tilt-lat}
\end{figure}

Obviously, this interpretation does not explain the behaviour of the small
bipoles. In order to clarify the situation we present in Figure~\ref{tilt-lat}
the behaviour of the tilt for small bipoles averaged in various latitudinal 
zones \textit{versus} time, and the distribution of the small bipoles compared with 
that of the large. We see from this figure that the small bipoles
demonstrate some kind of cyclic behaviour which is, however, 
quite different from that 
of the large. The time-latitude distribution of small bipoles demonstrates an
equatorward propagating pattern as well as poleward.
The cycle described by small bipoles looks shifted from that of 
the large. The tilt angles of the small bipoles are, as expected, 
determined mainly by bipoles located in the middle latitudes. 
In general, it looks plausible
that small bipoles represent a different component of the solar magnetic field 
to that traced by the large ones.

\subsection{Violations of Hale's Law}

Now we examine how Hale's polarity law works for large and small bipoles. 
Of course, there is a small fraction of bipoles which violate Hale's 
law. It is natural to compare this fraction with the fraction of sunspot 
groups which violate the law (according to \inlinecite{SK10}
this fraction is about  $5-7\%$).

\begin{figure}[h]
\begin{center}
\includegraphics[width=0.6\linewidth]{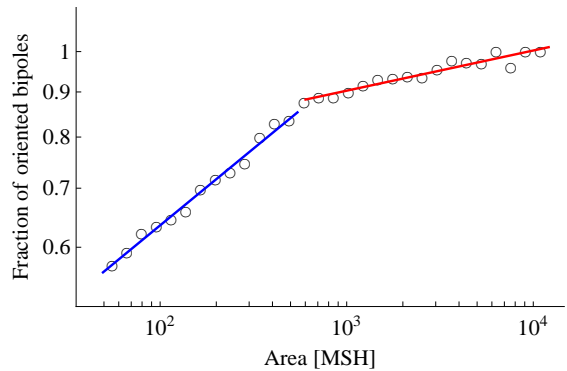} 
\end{center}
\caption{The fraction of bipoles oriented according to Hale's polarity law
(MDI data). Blue marks the segment with the slope of the line 0.2, red marks
the segment with the slope 0.05. }
\label{hale}
\end{figure}

The fraction of bipoles which {\it follow} the Hale polarity law \textit{versus} 
the bipole area is presented in Figure~\ref{hale}. The plot is organized as follows. 
We divide the bipole sample in bins according to their area. 
Dots in the plot correspond to centres of a bin. Then the fraction of bipoles 
in a bin which follow Hale's law is shown by the vertical coordinate 
in the plot. In our analysis we used the MDI data
at the maximum stage of the Cycle 23 (period 1998--2007). 
At the end of this cycle
the overlap of bipoles with opposed orientations occurs
because of the extended solar cycle at high latitudes 
\cite{T10}.

For the largest bipoles the fraction which 
follow Hale's law exceeds 90\%. Such bipoles correspond to bipolar 
sunspot groups and the result, as  expected, agrees with the estimate of \inlinecite{SK10}.  

The fraction of bipoles which follow Hale's law drops with the bipole area. 
This seems natural because it is more difficult to isolate small bipoles 
than the large ones, and the noise level in determination of the bipole 
orientation is larger for the small bipoles.  
The point however is that the plot in Figure~\ref{hale} shows specific slopes
for large, $S^{0.05}$, and small, $S^{0.2}$, bipoles. This confirms that small 
and large bipoles represent physical entities of different natures. The slopes
match near $S=500$ MHS.  This confirms the validity of the area threshold 
chosen to separate small and large bipoles. The data for bipoles in 
these groups are shown in blue and red in Figure~\ref{hale}. The fraction of 
bipoles which follow the Hale's law becomes as small as 50\% near $S=50$ MHS 
and, then, it becomes fruitless to consider smaller bipoles. 

Thus the estimate $5-7\%$ for the number of reversed bipoles is valid mainly
for bipoles with areas greater than 1000 MSH.

\subsection{Flux and Tilt}

\begin{figure}[h]
\begin{center}
\includegraphics[width=0.6\linewidth]{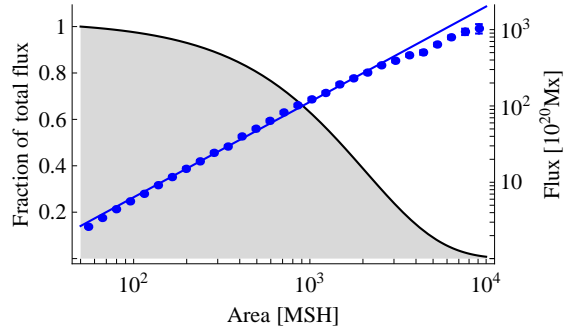} 
\end{center}
\caption{Black line shows the contribution of bipoles to the total 
flux according to MDI data (graphics of $\Phi(S)$). 
Blue dots indicate the mean flux for bipoles 
with different areas, the fitted line has a slope of 1.25.}
\label{sf}
\end{figure}

We now investigate in more detail the link between the size of 
a bipole and the tilt. There are two natural measures for the size of a bipole, 
its area $S$ and its magnetic flux $F$ (here and below $F$ is measured in 
$10^{20}$\,Mx). Fortunately, both these quantities are closely interrelated 
(Figure~\ref{sf}), $F(S)\sim S^{1.25}$ (blue line). 
This means that it is sufficient to study only the dependence on $F$. 
In the same figure we show that the large bipoles give the main contribution 
to the total magnetic flux. More precisely we plot with a black line
a function $\Phi(S)$ which gives the contribution to the total flux of bipoles 
with areas greater than $S$. The line is fitted well by 
$\Phi(S)=\exp [-S/(2\times 10^3)]$. 

Figure~\ref{sf} shows that about $90\%$ of total flux comes 
from bipoles with areas $S>300$ MSH. However we appreciate that such an
estimate does not, for instance, take into account that the lifetime of 
small bipoles (ephemeral regions) is shorter than that of the large 
bipoles (sunspots). Thus the contribution of the smaller bipoles to
the total flux may be underestimated and a more detailed analysis is required. 
In particular, a measure of the flux regeneration rate would seem 
to be more suitable here. 
   
\begin{figure}[h]
\begin{center}
\includegraphics[width=0.6\linewidth]{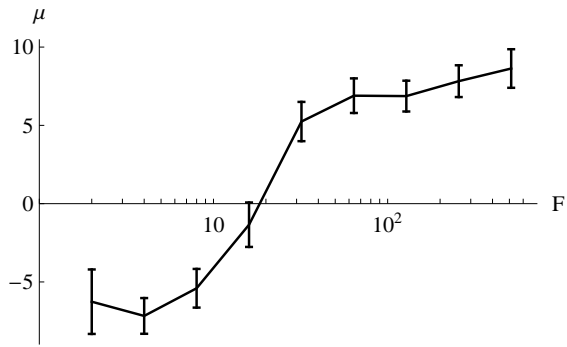}
\end{center}
\caption{Tilt against flux $F$ [$10^{20}$ Mx] for MDI data.}
\label{fl}
\end{figure}

Figure~\ref{fl} shows that a negative tilt value is predominant for $F<10$. With $F>30$
it becomes positive. Comparing the plot with the
previous Figure~\ref{sf} we conclude that the dividing point $F=20$
corresponds to an area $S=300$ MSH. The tendency of increasing tilt seems 
to remain for larger values of $F$. 

The investigation of the correlation between flux and tilt is interesting
in the first place for estimating the contribution of bipolar moments 
to the formation of the poloidal component of magnetic field \cite{S13}. 
We recall that the bipolar moment is $B_m=F\cdot d$, where $F$ is the flux of a
bipole 
and $d$ is the distance between the unipolar regions in the bipole. Furthermore, 
$d$ is another  natural measure of the bipole size.

The distance $d$ is defined as the distance in 
heliographic degrees between the geometrical centres of monopoles in 
a given bipole. This definition  includes indirectly a contribution from 
the area of the bipole (a part of $d$ comes from the radii of the two opposite polarities of the bipole)
and is strongly affected by the shape of the domains (complex configurations
can even give zero $d$). Indeed, a simplistic presentation of bipoles
as two near circular domains leads to $d$ increasing as $\sqrt{S}$,
where $S$ is a measure of area of a bipole.

\begin{figure}[h]
\begin{center}
\includegraphics[width=0.6\linewidth]{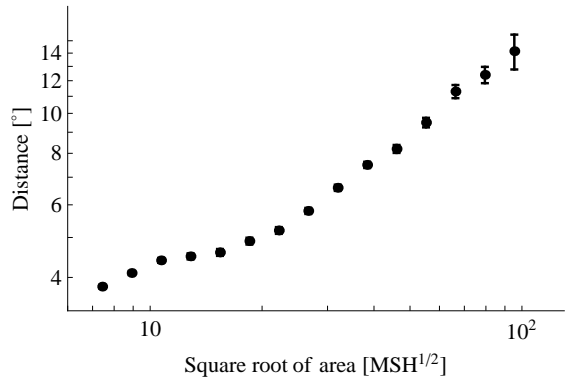} 
\end{center}
\caption{Mean distance $d$ against size of the domains given by $\sqrt{S}$.}
\label{sd}
\end{figure}
 
In fact Figure~\ref{sd} shows the slope of $d$ as function of $\sqrt{S}$ 
to be significantly less than one. This means that  domain sizes
increase faster than the distance between them. Again, behaviour of the 
plot is different for large and small bipoles (however the slope is 
almost the same), and the dividing point is close to $S=300$ MHS 
(note that the plot shows $\sqrt S$ rather than $S$).

\begin{figure}[h]
\begin{center}
\includegraphics[width=0.6\linewidth]{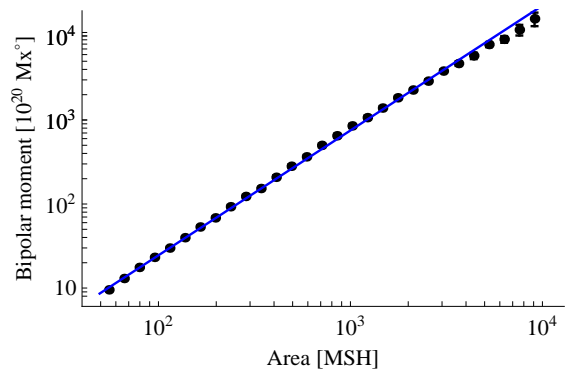}
\end{center} 
\caption{Bipolar moment against area of bipoles, MDI data.
The fitted line has a slope of 1.5.}
\label{sbm}
\end{figure}

\begin{figure}[h]
\begin{center}
\includegraphics[width=0.6\linewidth]{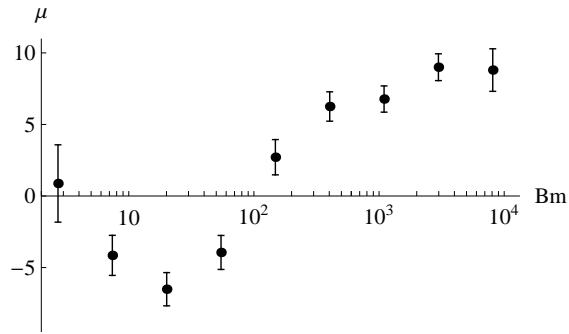}
\end{center}
\caption{Tilt against bipolar moment $Bm$ [$10^{20}$ Mx$^\circ$] for MDI data.} 
\label{tbm}
\end{figure}

For investigation of the bipolar moment, $Bm$, we consider first its 
dependence on the area of the bipoles. Figure~\ref{sbm} shows that $Bm$ 
is proportional to $S^{1.5}$. The mean tilt versus $Bm$ is shown in 
Figure~\ref{tbm}. We see that the tilt of small bipoles behaves again 
in the opposite way to that of the large bipoles and the dividing point 
($Bm=90$) lies between $200-300$ MSH (Figure~\ref{sbm}).

Figure~\ref{tbm} shows a moderate growth of the tilt with increasing bipolar 
moment for large bipoles. Note that \inlinecite{sk12} did not find any 
significant variations of tilt angle with flux or bipolar moment.

Now we can calculate the averaged effect of the tilt as follows. 
We consider the bipoles which follow Hale's polarity law, multiply 
the tilt by the polarity $p=\pm1$ of the leading component of the 
bipole, and sum $F\sin(p\mu)$ over all bipoles (for the period 
1998--2007 using MDI data). The quantity obtained 
is 10\% of the total magnetic flux and shows which part of the toroidal 
magnetic field is converted to poloidal one. In other words, it is an 
estimate of the ratio $\alpha/v$, where $v$ is the r.m.s. velocity 
of the convection and $\alpha$ is the magnitude of the alpha-effect. 
The estimate is robust in sense that it remains stable if the large 
bipoles only are taken into account (small bipoles give only a minor 
contribution to the estimate), or if we consider bipolar moments instead 
of magnetic fluxes. The estimate is remarkably close to the order-of-magnitude
estimate in \inlinecite{TIS} and corresponds to a traditional expectation 
from dynamo theory.

\section{Discussion and Conclusions}

Summarizing, we conclude that we can recognize specific properties of small 
and large bipoles, which are represented mainly by ephemeral regions and 
sunspot groups respectively. The dividing point between the groups is 
located near a bipole area $S = 300$ MSH. However the separation of the groups 
can be based on other relevant quantities, \textit{i.e.} bipolar moment, magnetic flux,
distance between the bipole domains. All ways of separating the 
groups give similar results. The main difference between the 
two groups in the context of our research is the opposite sign of the
bipole tilt.  However specific properties of two groups of 
bipoles are visible in other respects as well. 

We note several other remarkable features in the results obtained.  

The time-latitude diagram for small bipoles (Figure~\ref{tilt-lat}) seems 
to agree with the concept of the extended solar cycle \cite{Wetal88}. 
In particular the wings of the butterfly diagrams for the tilt of small 
bipoles start  at high  latitudes 1 -- 2 years earlier than the corresponding 
sunspot cycle, and then propagate towards the solar equator.  

For the large bipoles ($S > 300$ MSH), the tilt becomes larger for 
larger bipoles. The tendency is visible if the flux 
is considered as a measure of the size of bipoles (Figure~\ref{fl}) 
as well as if the bipole moment is considered as the 
measure (Figure~\ref{tbm}). 

Separation of bipoles into small and large with opposed 
properties seems to support  the idea of \inlinecite{Ch12} that the 
regimes of dynamo action in the presence  of sunspots (\textit{i.e.} large bipoles) 
and in their absence are substantially different. Probably, this 
illuminates a difference between solar dynamo action during the Maunder 
minimum and in contemporary solar cycles. Of course, a simple 
explanation is that the fields responsible for the large and small bipoles 
just originate from different depths where the fields have different 
properties.

Pragmatically, the contribution of large bipoles to the 
total $\alpha$-effect (that parametrizes solar dynamo action) dominates 
and it seems possible to ignore small bipoles when quantifing it. 
A deeper understanding of the processes underlying solar dynamo action 
deserves however a physical interpretation of the opposed properties of 
the large and small bipoles with respect to the tilt.

First of all, we have to note that our research is inevitably based on 
a complicated algorithmic procedure leading to the isolation of bipoles
from magnetograms. Our analysis does not show any trace of a bug
in the algorithm used which could produce a difference between small 
and large bipoles. It is difficult however to exclude such an option 
completely based on the results of the application of just one algorithm 
for isolation of the bipoles. We thus stress the desirability of comparing 
our results with those from other algorithms which can isolate small 
bipoles from large. Future research in this direction is needed.

\inlinecite{TIS} suggested that the opposite sign of tilt for small and large 
bipoles can be understood as indicating that large bipoles represent 
toroidal magnetic field while the small ones are connected with poloidal 
magnetic field. Figure~\ref{tilt-lat} seems to support this interpretation. 
Indeed, the interpretation of large bipoles, \textit{i.e.} sunspot groups as tracers 
of toroidal magnetic field, is standard and the problem is to identify what 
is traced by the small bipoles.
Figure~\ref{tilt-lat} shows that the tilt patterns in the time-latitude
diagram for small bipoles are pronouncedly dissimilar to the sunspot wings 
in the diagram. They are however quite similar to corresponding patterns 
of the large-scale surface magnetic field \cite{Oetal06} which presumably 
trace the solar poloidal magnetic field. The fact that we see specific slopes 
for small and large bipoles in other plots presented above seems to agree with 
this interpretation. However one could expect even more dramatic events at 
transitions in the plots between small and large bipoles, such as jumps. 
The point is that dynamo modelling predicts that toroidal magnetic field 
should be substantially stronger than the poloidal, which would be expected 
to result in jumps.

Another possible interpretation might be based on the idea that the 
$\alpha$-effect is associated with hydrodynamic and magnetic helicities, 
which are inviscid invariants of motion (\opencite{S96}; \opencite{K98}). 
Then accumulating helicity in one range of scales (or spatial region) 
would have to be compensated by growth of helicity of the opposite sign 
in the other range. A moderate growth of the tilt with bipolar moment 
for large bipoles in Figure~\ref{tbm} seems to be consistent with 
the idea of the helicity separation in Fourier space \cite{S96}: the point 
is that the Coriolis force is larger for larger bipoles. A continuous behaviour 
of the plots in figures which do not involve tilt directly becomes natural with
this interpretation: all bipoles now represent toroidal field.  
In contrast, the form of the time-latitude diagrams in Figure~\ref{tilt-lat}
now requires an explanation.

We stress that the arguments in favour of either of the above interpretations 
do not seem  convincing enough for us to prefer one to the other.

\begin{acks}[Acknowledgments]
The paper is supported by RFBR under grants 
12-02-00170,
13-02-91158,
12-02-00884,
12-02-31128,
13-02-01183,
12-02-00614.
We are grateful to D.~Moss for a critical reading of the paper.  
\end{acks}

\end{article}
\end{document}